\documentclass[iop]{emulateapj}
 \usepackage{amsmath, amsthm, amssymb,amsfonts}
\usepackage[english]{babel}
\usepackage[]{color}

\begin{document}

\newcommand{\va}{v_{\textsc a}}
\newcommand{\taua}{\tau_{\textsc a}}

\title{The tearing mode instability of thin current sheets: the transition to fast reconnection in the presence of viscosity}

\author{Anna Tenerani}
\affiliation{Jet Propulsion Laboratory, California Institute of Technology, Pasadena, CA}
\email{Presently at UCLA: annatenerani@epss.ucla.edu}
\author{Antonio Franco Rappazzo}
\affiliation{Advanced Heliophysics, Pasadena, CA}
\author{Marco Velli}
\affiliation{EPSS, UCLA, Los Angeles, CA}
\and
\author{Fulvia Pucci}
\affiliation{Universit\`a degli studi di Roma Tor Vergata, Rome, Italy}

\begin{abstract}
This paper studies the growth rate of reconnection instabilities in thin current sheets in the presence of both resistivity and viscosity. In a previous paper, \citet{pucci_2014}, it was argued that at sufficiently high Lundquist number $S$ it is impossible to form current sheets with aspect ratios $L/a$ which scale as $L/a \sim S^{\alpha}$ with $\alpha >1/3$ because
the growth rate of the tearing mode would then diverge in the ideal limit $S\rightarrow \infty$. Here we extend their analysis to include the effects of viscosity, (always present in numerical
simulations along with resistivity) and which may play a role in the solar corona and other astrophysical environments. A finite Prandtl number allows current sheets to reach larger aspect ratios
before becoming rapidly unstable in pile-up type regimes. Scalings with Lundquist and Prandtl numbers are discussed as well as the transition to kinetic reconnection.

\end{abstract}

\maketitle

\section{Introduction}

Current sheets are generically unstable to resistive reconnecting instabilities, the archetype of which is the tearing mode~\citep{FKR}. In the tearing mode instability
the fastest growing perturbations have wavelengths along the current sheet which are much greater than the thickness of the sheet itself, identified with the shear scale-length of the equilibrium magnetic field. Magnetic islands develop in a small region around  the neutral line  (or, more generally, for a current sheet with a non-vanishing axial field, on surfaces where the wave vector is such that  ${\bf k\cdot B}_0=0$) and grow on time scales intermediate between the diffusion time scale of the equilibrium and the Alfv\'en time scale as measured on the current sheet thickness.

Though the tearing mode has been studied extensively over the past decades, its role in triggering fast magnetic energy release and current sheet disruption has attracted recent research 
focusing on the stability of thin current sheets, starting from the Sweet-Parker steady state reconnecting configuration (which is found to be unstable to the plasmoid instability~\citep{lou_2007, lou_2012}), and on the role of kinetic effects in speeding up reconnection compared to the slow growth rates originally found in~\citep{FKR}. 

It was recently suggested in Ref.~\citep{pucci_2014} that the use of the Sweet-Parker current sheet as potential initial configuration leading to fast instability was misleading in the limit of very large Lundquist numbers, because of the fast plasmoid instability whose growth rate  diverges as the Lundquist number, based on the macroscopic scale, tends to infinity. In that paper it was shown that a current sheet with a limiting aspect ratio much smaller than that of the Sweet-Parker sheet, scaling as $S^{1/3}$ ($S$ being the Lundquist number), separates slowly growing resistively reconnecting sheets from those exhibiting fast plasmoid instabilities, and that this provides the proper convergence properties to ideal MHD, which is a singular limit of the resistive MHD equations. In Ref.~\citep{pucci_2014},  the tearing mode instability is studied for a family of current sheets with aspect ratios scaling as $L/a=S^\alpha$. Interestingly, if $\alpha <1/2$, the Sweet-Parker value, static equilibria may be constructed that do not diffuse, while flows in the Sweet-Parker model are required, as the current sheet would otherwise diffuse on an ideal time-scale based on the macroscopic current sheet length. \citet{pucci_2014}  find that as $\alpha \rightarrow 1/3$ from below, the growth rate becomes independent of $S$ itself, reaching a value of order unity. As such, that aspect ratio provides a physical upper limit to  current sheet aspect ratios that may form naturally in plasmas. Otherwise, at large $S$, the instability time-scale would become faster than the time required to set-up the equilibrium in the first place.

The linear study of reconnection instabilities as described above is intended as a useful schematization to inspect the local evolution of quasi singular current layers which, in fact, naturally form when embedded in a far richer and dynamical context such as, for example, in the field line-tangling of the Parker nanoflare model of coronal heating~\citep{rappa_2013}, or the propagation of waves around x-points or along separatrix surfaces in 3D fields, e.g., at the boundaries of closed and open fields in the so-called streamer and pseudo-streamer configurations in the solar corona. 

Therefore, the question arises as to which other effects limit the aspect ratios of current sheets that can be formed  before they disrupt on ideal time scales. Also,  given the extremely large aspect ratios of  \textquotedblleft ideally\textquotedblright\ reconnecting current sheets,  microscopic processes, such as two-fluid and kinetic effects, might be  fundamental in the initial stages of the fast current disruption~\citep{cassak_prl_2005, cassak_pop_2013}.     

In this paper we focus on the effects of viscosity in determining limiting values for current sheet dimensions, i.e., in determining at which aspect ratios  the growth rate of reconnection instabilities reaches ideal values --assuming a scaling of the aspect ratio with  the macroscopic Lundquist  and Prandtl number. In the  magnetized limit  $\omega_{ci(e)}\tau_{i(e)}\gg1$, $\omega_{ci(e)}$ and $\tau_{i(e)}$ being the ion (electron) Larmor frequency and collision time, respectively, there are two different viscous  transport coefficients of the plasma, along the direction parallel and perpendicular  to the mean magnetic field~\citep{brag}. Indicating with $\nu$ and $\nu_\parallel$  perpendicular and parallel (ion) kinematic viscosity, and with $\eta$ the parallel magnetic diffusivity, the Prandtl numbers are given by 
\begin{equation}
P\equiv\frac{\nu}{\eta}=0.3\frac{4\pi}{\sqrt{2}}\frac{ n k T_i}{B^2}\sqrt{\frac{m_i}{m_e}}\left(\frac{T_e}{T_i}\right)^{3/2}\approx\beta\sqrt{\frac{m_i}{m_e}},
\end{equation}
\begin{equation}
P_\parallel\equiv\frac{\nu_\parallel}{\eta}=0.96\frac{4\pi}{c^2}\frac{n e^2\tau_e}{m_e}\frac{k T_i\tau_i }{m_i}\approx5\times10^{-6}\frac{T^4}{n},
\end{equation}
where $\beta$ is the ratio of ion thermal pressure to magnetic pressure, $n$ the number density, and $T_{i,e}\approx T$ are the ion and electron temperatures, respectively.  As a consequence, in many astrophysical environments the Prandtl number may range from small values to values of order one or larger~\citep{dobro, sheko_2005}:   in table~\ref{table1} we list for reference some examples of magnetized plasmas that can be found in space, and their order of magnitude parameters. It is therefore of interest to study viscous effects on the stability of thin current sheets over a broad range of Prandtl numbers.

We will begin in Section~\ref{uno} by introducing the basic set of equations describing the tearing mode and by clarifying the notation used throughout the paper. Section~\ref{classic} and~\ref{para} follow with a detailed description of the \textquotedblleft classic\textquotedblright\ visco-resistive tearing instability (i.e., current sheets of a given thickness), providing a unified framework summarizing previous results scattered in the literature and setting the stage for the subsequent focus on the scalings of growth rates and singular layer thickness with Lundquist and Prandtl (or magnetic Reynolds) numbers for arbitrary aspect ratios. The latter are discussed in Section~\ref{collapsing}, where we also  consider the possible consequences of the onset of reconnection at high Prandtl numbers. Conclusions are in Section~\ref{conclusions}.

\section{Model  equations}
\label{uno}

We consider the visco-resistive tearing instability of a current sheet within the framework of incompressible MHD.  In general, viscous forces in a magnetized plasma are described by the divergence of a stress tensor which involves both the parallel and perpendicular viscous coefficients  $\nu$ and $\nu_\parallel$~\citep{brag}. 
Moreover, the complete stress tensor contains also non-dissipative finite Larmor radius terms, sometimes called \emph{gyroviscosity} terms,  proportional to $nT_i/\omega_{ci}$~\citep{brag, cerri_2013}. We  will neglect them here,  with the reminder that in future generalizations to kinetic reconnection regimes gyroviscosity must be taken into account.

We schematize the current sheet with a sheared magnetic field ${\bf B}_0$ in slab geometry, ${\bf B}_0={\bf \hat y}\bar B_0\tanh{(x/a})$, where $a$ is its  width, and we assume a uniform mass density  $\rho_0$. The gradient of the magnetic pressure may be balanced either by a pressure gradient or by an inhomogeneous component of the guide field, along the $z$ axis, which may vary in such a way as to maintain a force-free configuration. 

The resulting set of incompressible MHD equations for a 2D geometry in the limit of strong guide field is
\begin{subequations}
\begin{equation}
\frac{\partial{\bf u}}{\partial t}+{\bf u\cdot\boldsymbol\nabla u}=-\frac{1}{\rho}{\boldsymbol\nabla}\left( p+\frac{ B^2}{2 }\right)+\frac{1}{\rho}{\bf B\cdot\boldsymbol\nabla B}+\nu\nabla^2{\bf u},
\end{equation}
\begin{equation}
\frac{\partial{\bf B}}{\partial t}={\bf\boldsymbol\nabla\times( u\times B)}+\eta\nabla^2{\bf B},
\end{equation}
\label{strong}
\end{subequations}
where ${\bf u}$ and ${\bf B}$ are the plasma velocity and magnetic field. In the opposite limit of weak guide field we get a non isotropic contribution of viscosity:
\begin{subequations}
\begin{equation}
\begin{split}
\frac{\partial{\bf u}}{\partial t}&+{\bf u\cdot\boldsymbol\nabla u}=-\frac{1}{\rho}{\boldsymbol\nabla}\left( p+\frac{ B^2}{2 }\right)+\frac{1}{\rho}{\bf B\cdot\boldsymbol\nabla B} \\
&+\left[\nu_\parallel \frac{\partial^2 u_x}{\partial x^2}  +\nu\frac{\partial^2 u_x}{\partial x^2}+4\nu\left(\frac{\partial^2 u_y}{\partial x\partial y}+\frac{\partial^2 u_x}{\partial y^2}\right)\right]{\bf \hat x}\\
&+\left[2\nu_\parallel\frac{\partial^2 u_y}{\partial y^2}+4\nu\left(\frac{\partial^2 u_x}{\partial x\partial y}+\frac{\partial^2 u_y}{\partial x^2}\right)\right]{\bf \hat y},
\end{split}
\end{equation}
\begin{equation}
\frac{\partial{\bf B}}{\partial t}={\bf\boldsymbol\nabla\times(u\times B)}+\eta\nabla^2{\bf B}.
\end{equation}
\label{weak}
\end{subequations}

Hereafter we will limit our analysis mainly to the effects of perpendicular kinematic viscosity $\nu$. Therefore, unless specified,  we  will consider  equations~(\ref{strong}), which are a valid approximation both in the RMHD ordering with strong  guide magnetic field and in the case of a weak guide field, since the transverse gradient is dominant  in the reconnection layer. We will discuss in Section~\ref{para} some of the effects of large parallel viscosity, and in which limit the set of equations~(\ref{strong}) may approximate equations~(\ref{weak}).

 We introduce in the equations to be discussed below a macroscopic length scale $ L$~\citep{pucci_2014} which represents the relevant spatial scale of the system, e.g., the length  of the sheet, with respect to which we define the three time scales of the system: the ideal Alfv\'en ($\taua$), the diffusive ($\tau_{\nu}$), and the resistive  ($\tau_{\eta}$) time scale,
\begin{equation}
\taua=\frac{L}{\va},\quad \frac{\tau_{\nu}}{\taua}=\frac{L\va}{\nu}\equiv R,\quad \frac{\tau_{\eta}}{\taua}=\frac{L\va}{\eta}\equiv S,
\label{norm}
\end{equation}
where $\va=\bar B_0/\sqrt{\rho_0}$ is the Alfv\'en speed. Following the usual notation, we have introduced the Lundquist number $S$, and  we labeled the kinematic Reynolds number (defined using the Alfv\'en velocity) with $R$. 

For the sake of clarity, we will refer throughout the  text  to the \textquotedblleft classic\textquotedblright\ tearing instability when time scales are measured with respect to the shear length $a$, which is achieved by setting $L=a$.  

By assuming that small perturbations are functions of the form $ f(x/a)\exp(i k y +\gamma t)$, $k$ and $\gamma$ being the wave vector  and growth rate of a given mode, respectively, linearization of the parent system of equations around the prescribed equilibrium leads to  two coupled equations for the (normalized) velocity  and magnetic field perturbations $\hat u$ and $\hat b$:
\begin{equation}
\begin{split}
\gamma\taua\frac{a^2}{L^2}&(\hat{u}^{\prime\prime}-\hat k^2\hat{u})=-\frac{\hat k}{\hat \rho_0}[\hat B_0(\hat b^{\prime\prime}-\hat k^2\hat b)-\hat b\hat B_0^{\prime\prime}]\\
&+R^{-1}[(\hat u^{\text{\sc iv}}-\hat k^2\hat u^{\prime\prime})-\hat k^2(\hat u^{\prime\prime}-\hat k^2\hat u)],
\label{mom}
\end{split}
\end{equation}
\begin{equation}
\gamma\taua b=\hat u\hat k\hat B_0+S^{-1}\frac{L^2}{a^2}(\hat b^{\prime\prime}-\hat k^2\hat b).
\label{fara}
\end{equation}
In the  equations above, a {\it prime} denotes differentiation with respect to the normalized variable $x/a$, and $\hat k=ka$. Magnetic fields are normalized to $\bar B_0$, thus $\hat B_0=B_0/\bar B_0$ and $\hat b=b_x/\bar B_0$, and the normalized velocity is $\hat u=iu_xL/(\va a)$.  

There are a number of previous studies analyzing the effect of viscosity on \textquotedblleft classic\textquotedblright\ tearing modes. While exact solutions to equations~(\ref{mom})--(\ref{fara}) could not be found, approximated solutions in the constant-$\psi$ and non constant-$\psi$ (resistive internal kink) regimes~\citep{porcelli_1987,grasso_pop_2008, militello_pop_2011}, showed that even a moderate value of the viscosity has non negligible effects.  In particular,  viscosity becomes more important for increasing wave vectors, towards  marginal stability ($\gamma=0$): as is intuitive, viscosity, on the one hand tends to slow down the instability with respect to the inviscid case, and, on the other hand, it prevents the reconnective layer $\delta$ from shrinking indefinitely as marginal stability is approached, $\Delta^{\prime}\rightarrow0$ (or,  for our equilibrium, $ka\rightarrow1$)~\citep{bondeson, grasso_pop_2008, militello_pop_2011}.  \citet{porcelli_1987} provides the most interesting and relevant results concerning  visco-resistive tearing. He shows that the growth rate scales as $\gamma\taua\sim S^{-5/6}R^{-1/6}$ in the constant-$\psi$ regime, and $\gamma\taua\sim S^{-2/3}R^{-1/3}$ in the non constant-$\psi$ one, whereas the inner reconnective layer scales as $\delta\sim(SR)^{1/6}$ in both regimes. In addition, both analytical and numerical calculations confirmed that  viscosity removes the singularity at $\gamma=0$ and allows for the existence of  non-singular marginal  modes  at finite values of $\Delta^{\prime}$ (i.e., $ka<1$).  

In the next Section we carry out a more comprehensive analysis in parameter space, and we consider asymptotic scalings with $S$ and~$R$.

\section{Visco-resistive tearing mode}
\label{classic}
 
In this Section we describe our main numerical results on the \textquotedblleft classic\textquotedblright\ viscoresistive tearing instability, where, following the historical approach, the time scales $\taua$, $\tau_\nu$, and $\tau_\eta$  are defined via the shear length $a$. 

Equations~(\ref{mom})--(\ref{fara})~have been integrated numerically with an adaptive finite difference scheme, based on Newton iteration, which was designed by Lentini and Pereira in the seventies to  solve two-point boundary value problems for systems of ODE~\citep{lentini}.  The maximum absolute error on the solution is specified and the boundary layer structure of the solution is solved by increasing the mesh points in that region. This method has become a standard numerical technique supplementing the asymptotic analysis for linear plasma stability problems (see e.g., \citep{malara_96, Velli_1989}, and references therein).

\begin{figure}[htbp]
\begin{center}
\includegraphics[width=0.45\textwidth]{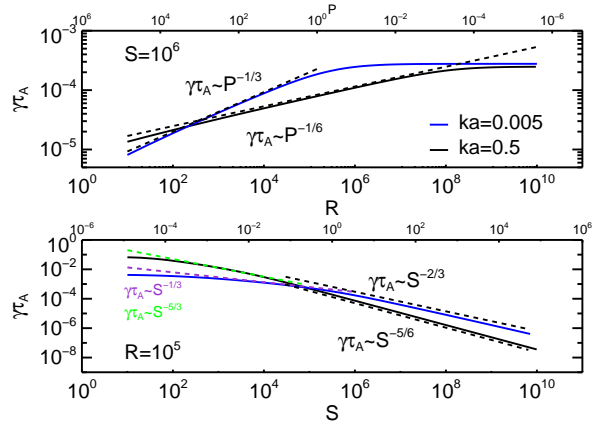}
\caption{Growth rate as a function of $R$  and of the Prandtl number  for ${S}=10^6$ (upper plot) and as a function of ${S}$ and $P$ for ${R}=10^5$ (lower plot)  for $\hat k=0.5$ and $\hat k=0.005$, which are above (const-$\psi$) and below (non const-$\psi$) the fastest growing mode, respectively. }
\label{constpsi}
\end{center}
\end{figure}

\begin{figure*}[htbp]
\begin{center}
\epsscale{0.5}
\plotone{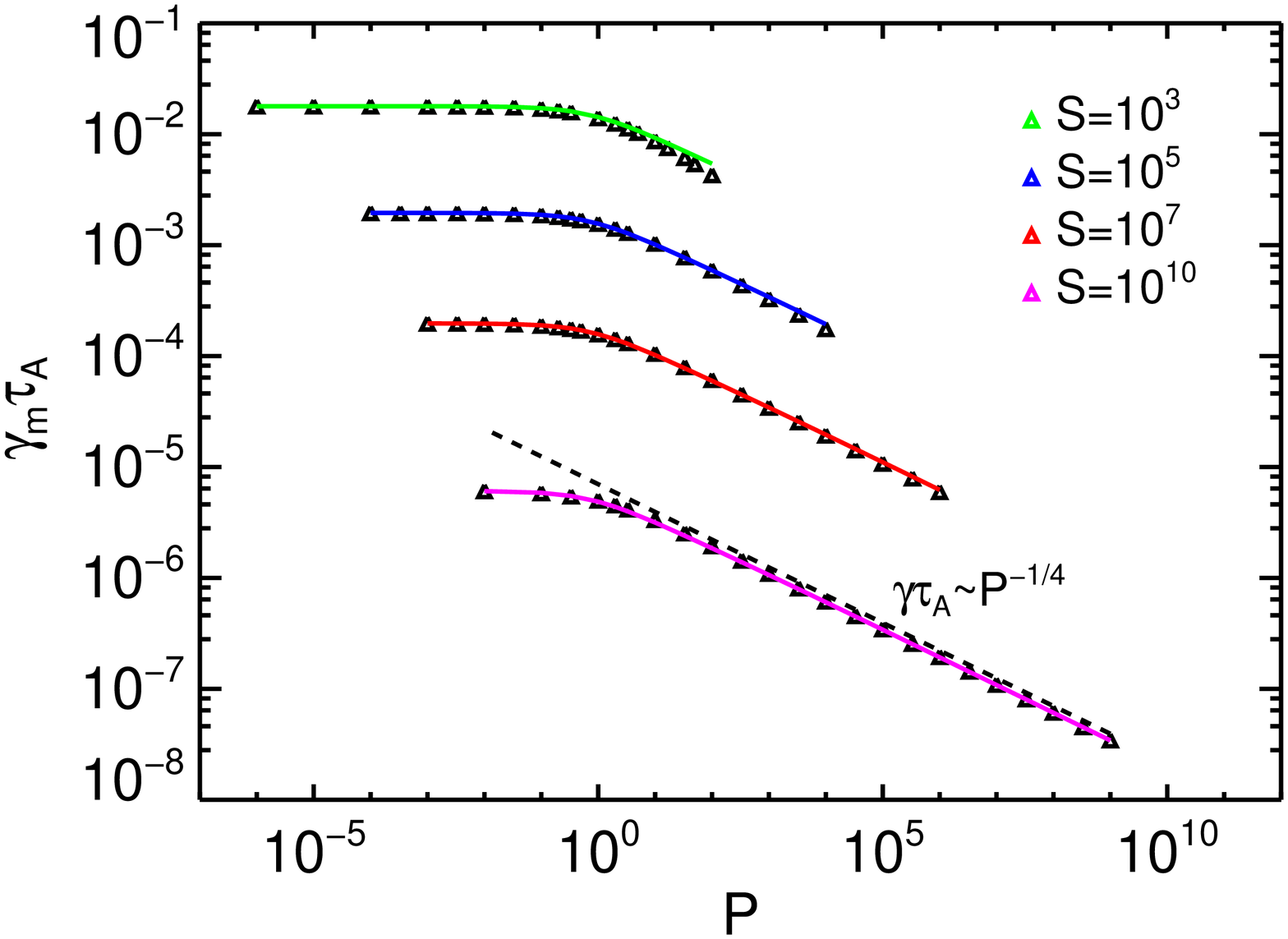}\qquad\qquad
\plotone{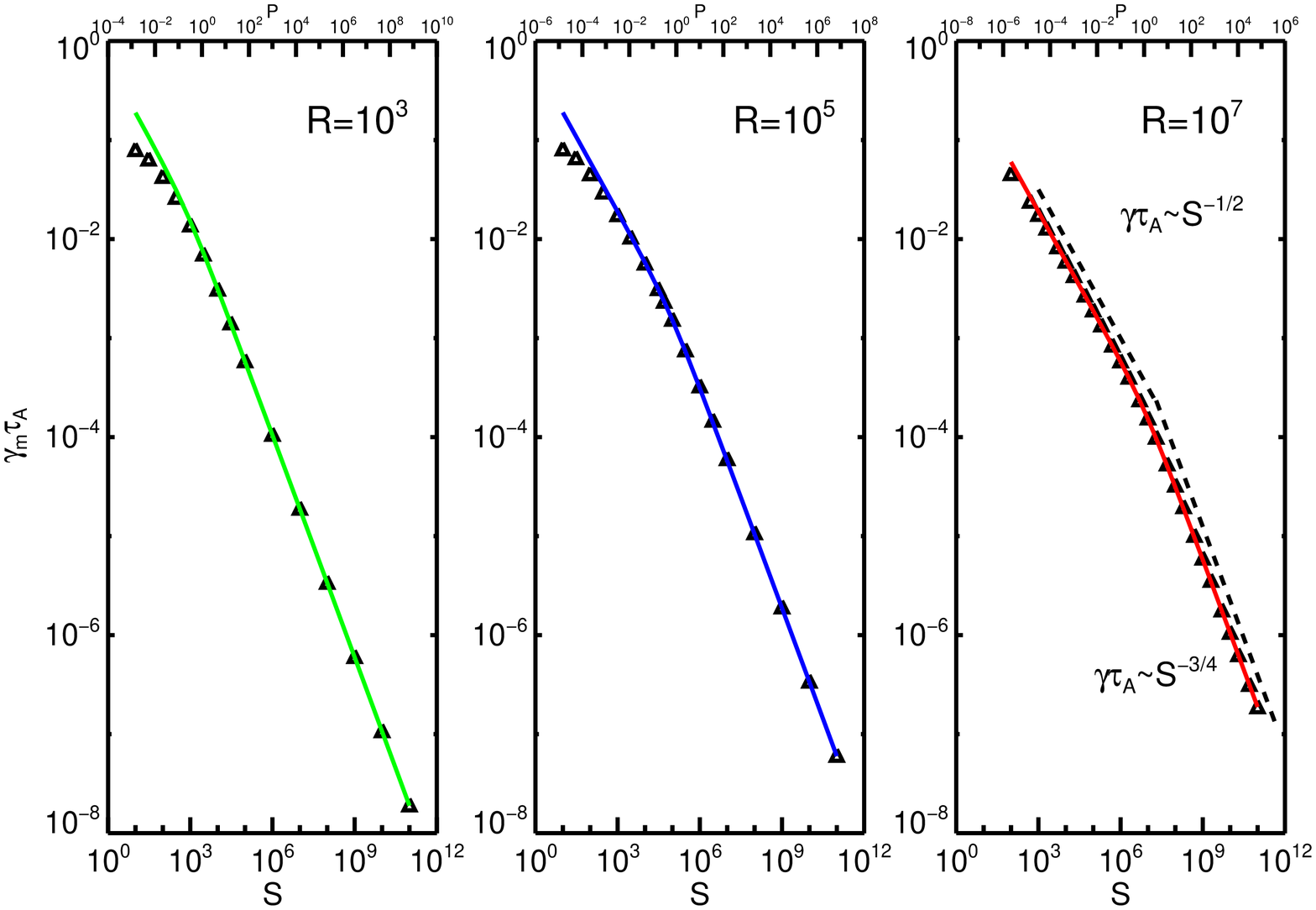}
\caption{Normalized growth rate of the fastest growing mode $\gamma_m\taua$. Left panel: $\gamma_m\taua$ versus  Prandtl number $P$  for four fixed values of Lundquist number $S$. Right panel: $\gamma_m\taua$ versus $S$, lower abscissa, and $P$, upper abscissa, for three different Reynolds numbers $R$. Colored lines which fit the numerical points are given by equation~(\ref{fit}). Black dashed lines correspond to the  scalings valid for the two  limit cases $P\gg 1$ and $P< 1$.}
\label{fgm}
\end{center}
\end{figure*}

\begin{figure*}[htbp]
\begin{center}
\includegraphics[width=0.45\textwidth]{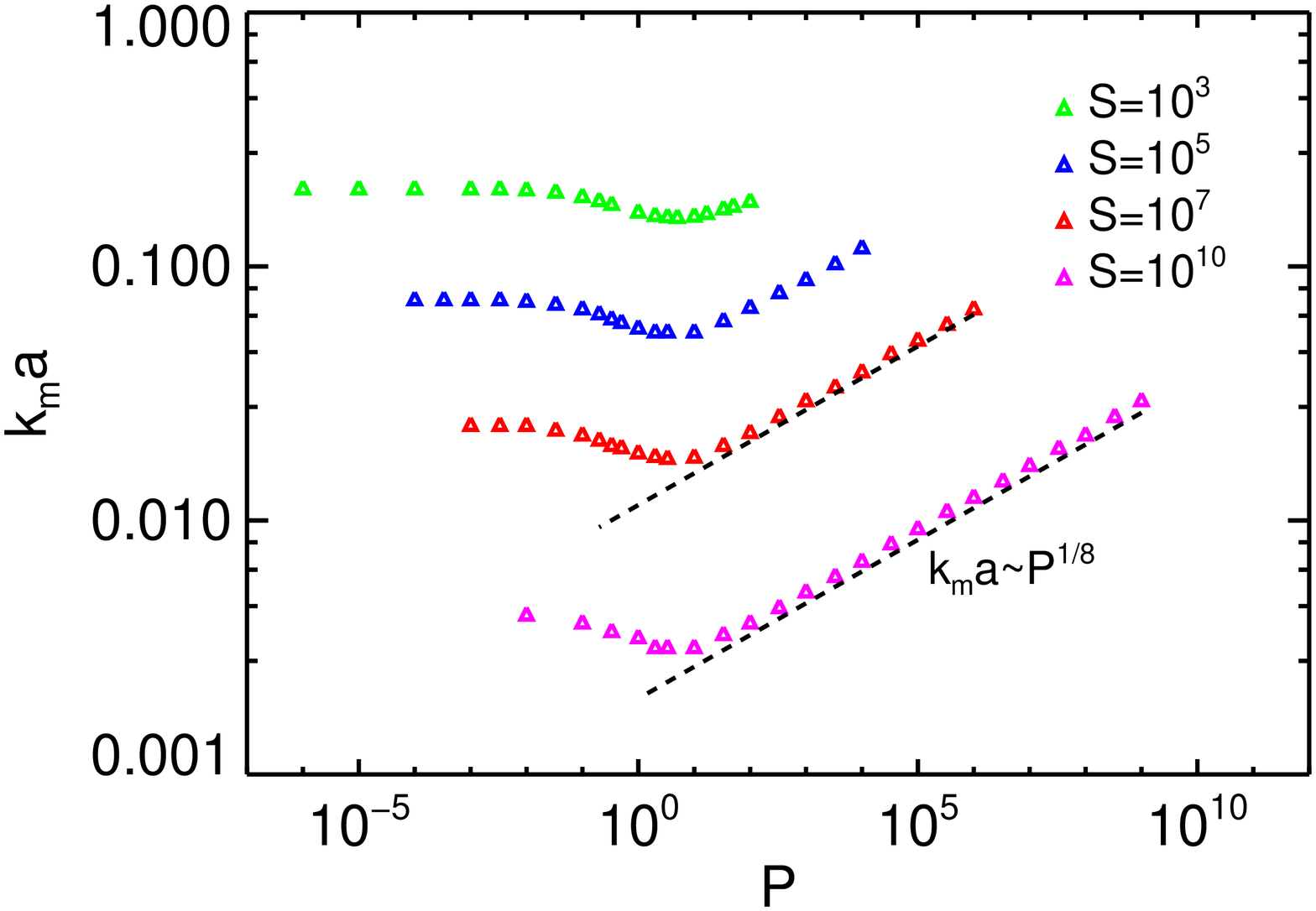}\qquad
\includegraphics[width=0.45\textwidth]{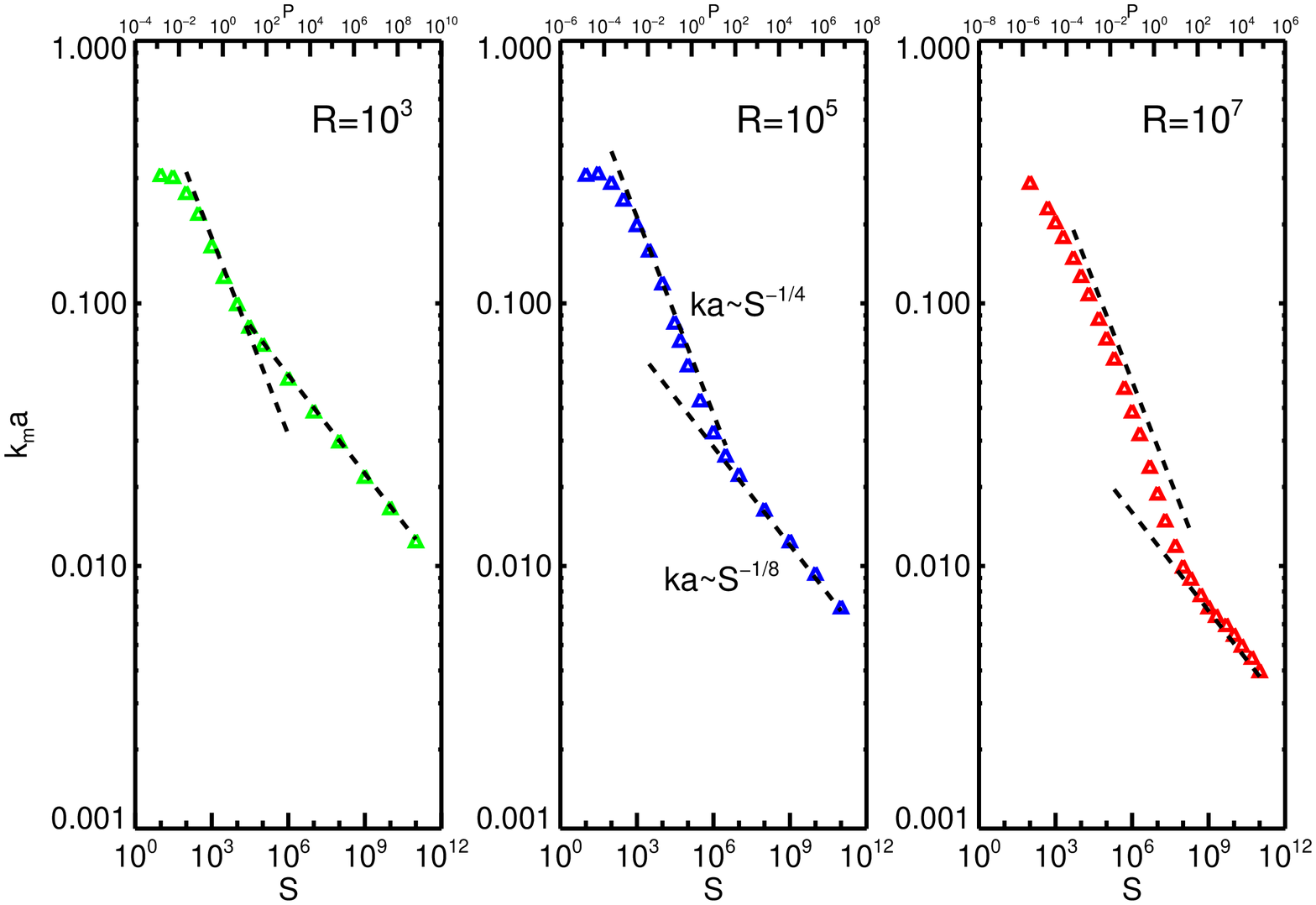}
\caption{Normalized wave vector of  the fastest growing mode $k_ma$ with the same parameters as in  Fig.~\ref{fgm}:  $k_ma$ versus $P$, in the left hand panel, and versus $S$ and $P$, right hand panel. Dashed lines represent the scalings valid for $P<1$ and $P\gg 1$.}
\label{k_fgm}
\end{center}
\end{figure*}

\begin{figure}[htbp]
\begin{center}
\includegraphics[width=0.45\textwidth]{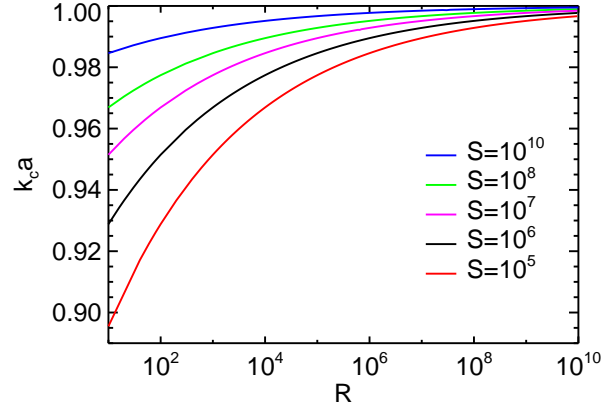}\qquad
\caption{Wave vector at marginal stability, $\gamma=0$, versus Reynolds number $R$ for different Lundquist numbers $S$. }
\label{kc}
\end{center}
\end{figure}

We integrated the eigenmode equations for a given wavevector $\hat k$  by imposing at the boundaries to the left and to the right of the magnetic neutral line where both viscosity and resistivity can be neglected, the outer layer solution of the tearing mode which goes  to zero for $|x|\rightarrow\infty$~\citep{Velli_1989}.  
Our results are summarized from Fig.~\ref{constpsi} through Fig.~\ref{kc}. We  recover the known analytical results, and we extend and complete the numerical analysis to a wider range of parameters. The Prandtl  number $P=S/R$ is allowed to vary from high values, $P\gg1$, all the way down to $P\ll1$, by changing either the Lundquist number $S$ at fixed  Reynolds number $R$ or, vice-versa, by varying $R$ at fixed $S$. The dependence on $S$ and $P$, or $S$ and $R$, of the growth rate of the fastest growing mode $\gamma_m$, and of the correspondent wave vector $k_m$,  is also described in detail. Our focus on the fastest growing mode stems from the idea that once a current sheet becomes increasingly thin, then one expects the fastest growing mode to dominate  the evolution of the instability. 

\begin{figure}[htbp]
\begin{center}
\includegraphics[width=0.45\textwidth]{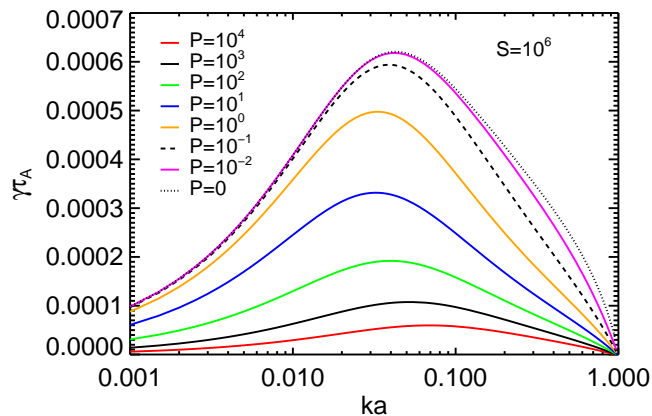}
\caption{Dispersion relation $\gamma\taua$ versus $ka$ for ${S}=10^6$ at different Prandtl numbers (or ${R}= \infty,\,10^8,\, 10^7,\, 10^6,\, 10^5,\,10^4,\,10^3,\,10^2$).}
\label{rd}
\end{center}
\end{figure}

In Fig.~\ref{rd},  we show an example of the dispersion relation  in the range $0.001\le \hat k \le1$  for $S=10^6$ and different Reynolds numbers, which correspond to Prandtl numbers   $10^{-2}\le P\le 10^4$. The inviscid case, in dotted line, is recovered asymptotically for $P\rightarrow0$. It can be seen that a small, but finite, viscosity affects modes with relatively large wave vectors,  about $\hat  k\lesssim1$: the growth rate is reduced and, as will be discussed  below, there exists a critical wave vector $k_c$ above which the equilibrium is  stable ($\gamma<0$). On the contrary, modes with smaller wave vectors, $\hat k\ll1$,  deviate from  their asymptotic values, defined at $P=0$, for  higher Prandtl numbers, as can be seen by comparing curves with $P\le 1$ vs. those with $P>1$. In Fig.~\ref{constpsi}, upper plot,  the growth rate for two chosen values of the wave vector is plotted as a function of $R$ (lower abscissa) and $P$ (upper abscissa),  and, in the lower plot, we show the growth rate, for the same modes,  as a function of $S$ and $P$. The two modes have wave vectors $\hat k=0.5$ and $\hat k=0.005$, which lye above and  below the fastest growing mode, respectively. Roughly speaking, the former corresponds to the constant-$\psi$ regime and the latter to the  non constant-$\psi$ regime. In both cases, the growth rate increases for decreasing viscosity, eventually becoming independent from viscosity itself, as can be seen from  the plateau which forms at $P\lesssim1$. It is clearly seen now that the mode with larger  wave vector reaches the plateau  for smaller values of viscosity ($P\ll1$). Before the plateau, the scaling valid in the constant-$\psi$ approximation $\gamma\taua\sim P^{-1/6}$~\citep{porcelli_1987, ofman} is recovered at intermediate values of viscosity $10^{-2}\lesssim P\lesssim10^2$ for $\hat k=0.5$, while the scaling $\gamma\taua\sim P^{-1/3}$ is obtained in the non constant-$\psi$ regime, $\hat k=0.005$~\citep{porcelli_1987}. Similarly,  the lower plot shows that for $P\ge1$ the growth rate scales as $\gamma\taua\sim S^{-5/6}$ for $\hat  k=0.5$ and $\gamma\taua\sim S^{-2/3}$ for $\hat k=0.005$ (black dashed lines).  For smaller Prandtl numbers, in an interval spanning from about $P>10^{-2}$ to  $P<10^{-1}-1$, the growth rate follows the two known scalings for the constant-$\psi$ and non constant-$\psi$ regime --provided $S$ is large enough, plotted for reference in green and violet dashed lines, respectively. 

With the intent of inferring the scalings of the fastest growing mode with Lundquist and Prandtl numbers  $S$ and~$P$ (or with the Reynolds number $R$), we plot  in Fig.~\ref{fgm} and~\ref{k_fgm} the maximum growth rate $\gamma_m$ and respective wave vector $k_m$ versus $P$  (left panels) and versus both $S$ and $P$ (right panels).  In the left hand panels we spanned from Reynolds  $R<S$ ($P>1$) to  $R>S$ ($P<1$) for fixed $S$. Similarly, in the right hand panels we chose three different values of $R$ and  spanned from $S<R$ to $S>R$. In the right hand plots we show in the lower abscissa  the Lundquist number and in the upper abscissa, for reference, the Prandtl number. By inspection of  numerical results displayed in Fig.~\ref{fgm},  we found an expression which represents  the maximum growth rate in the asymptotic limit $S\gg1$:
%
\begin{equation}
\gamma_m\taua=\left[ \frac{f(P)}{P+f(P)} \right]{\bar \gamma\taua}\footnote{The expression $f(P)=P+P^{3/4}+P^{1/2}+2$ fits the numerical solution to a very good approximation.},
\label{fit}
\end{equation}
where $f(P)\rightarrow1$ for $P\ll1$ and $f(P)\rightarrow P^{3/4}$ for $P\gg1$.  $\bar\gamma\taua\propto S^{-1/2}$ is the maximum growth rate in an inviscid plasma,  which is recovered by equation~(\ref{fit}) in the limit $P\ll1$. In the opposite limit $P\gg1$,  which is of major interest to us,  equation~(\ref{fit}) tends,  in agreement with~\citet{lou_2013}, to
\begin{equation}
\gamma_m\taua\sim S^{-1/2}\,P^{-1/4}=S^{-3/4}\,R^{1/4}.
\label{fit1}
\end{equation}

We used the expression~(\ref{fit}) to fit the numerical points in Fig.~\ref{fgm}, as  represented by  the superposed colored lines. In the left hand panel plot, we  solved the inviscid equations to determine the growth rate $\bar\gamma$, so as to find the exact asymptotic value reached by $\gamma_m\taua$  when approaching the inviscid limit $P\ll1$, which is achieved in practice at $P\approx0.1$. In the right hand panel plot, instead, the scaling $\bar\gamma\taua=cS^{-1/2}$ has been used,  and we chose an arbitrary constant to best fit the numerical points, which approaches the  value $c=0.62$ for increasing values of $R$. The dashed black lines are reported for reference and represent the scalings valid for $P\ll 1$ and  $P\gg 1$. It can be observed that the fit is increasingly accurate for higher values of the Lundquist number. 

A similar expression for the wave vector $k_m$ could not be found. Nevertheless, we inferred the scaling  $k_ma\propto S^{-1/8}\,R^{-1/8}$ for $P\gg1$, while the inviscid scaling $k_ma\propto S^{-1/4}$ is recovered for $P<1$. Black dashed lines are displayed to show the two scalings. Observe  that the wave vector $k_ma(R)$ has a minimum, as represented in the left hand panel. This can   be seen also  by inspection of Fig.~\ref{constpsi}, by following  the wave vector of the fastest growing mode which decreases with decreasing viscosity up to a minimum value and  increases again  (from the red to the magenta line).

Finally, as discussed  in the previous Section, viscosity allows for the existence of a marginally stable mode with $\gamma=0$. The critical wave vector $k_c$  separating modes with $k>k_c$ which are stable from those with $k<k_c$ which are unstable, is plotted in Fig.~\ref{kc} as a function of $R$ for different values of $S$.   In the limit $R\rightarrow\infty$, the marginal mode tends asymptotically  to $k_ca\rightarrow 1$, in agreement with the stability threshold condition for the inviscid tearing mode of a Harris current sheet.
As can be seen, while $k_c$ decreases for decreasing $R$ (increasing viscosity), as is intuitive, on the contrary,  for  fixed $R$,  the range of unstable modes becomes larger for increasing $S$ (decreasing resistivity).  Though the stabilization is weak, since for high Lundquist numbers $k_c$  is above  $k_ca\approx0.9$, it is interesting to remark that the marginal mode actually corresponds to a  configuration of {\it stationary} magnetic islands. This means that, at least in the linear approximation, the perturbed magnetic field ${\bf B_0+b}$ provides, in turn, an equilibrium where the current sheet is reconnecting. 

\begin{figure}[htbp]
\begin{center}
\includegraphics[width=0.45\textwidth]{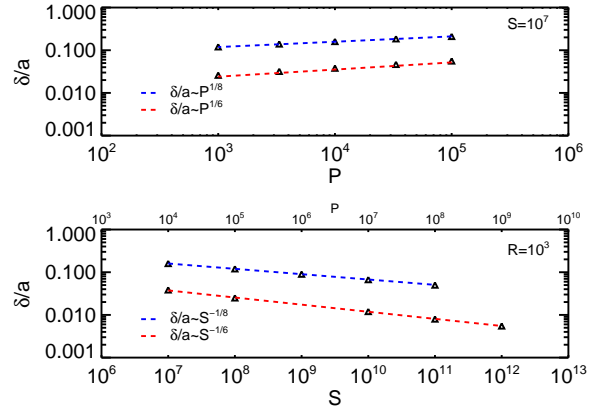}
\caption{Reconnective layer $\delta/a$ as a function of $P$ (upper plot) and $S$ (lower plot) at high Prandtl numbers. Points fitted with blue dashed lines correspond to the fastest growing mode and those fitted with red dashed lines correspond to the marginal mode.}
\label{delta}
\end{center}
\end{figure}

The width of the reconnective layer $\delta/a$ at high Prandtl numbers  as a function of $P$ and $S$ is plotted  in Fig.~\ref{delta},  fitted by the red dashed lines.  For comparison, we report also $\delta/a$ for the fastest growing mode, fitted by the blue dashed lines. The layer of the marginal mode scales as $\delta/a\sim(SR)^{-1/6}=P^{1/6}S^{-1/3}$, as found in the const-$\psi$ regime~\citep{porcelli_1987}. The layer of the fastest growing mode instead scales as $\delta/a\sim(SR)^{-1/8}=P^{1/8}S^{-1/6}$.

\section{Effects of parallel viscosity}
\label{para}

We discuss here the effects of large parallel Prandtl numbers $P_\parallel$ on the classic tearing mode instability. With obvious notation,  linearization of equations~(\ref{weak}) leads to
\begin{equation}
\begin{split}
\gamma\taua\frac{a^2}{L^2}&(\hat{u}^{\prime\prime}-\hat k^2\hat{u})=-\frac{\hat k}{\hat \rho_0}[\hat B_0(\hat b^{\prime\prime}-\hat k^2\hat b)-\hat b\hat B_0^{\prime\prime}]\\
&+4R^{-1}\hat u^{\text{\sc iv}}-3R_\parallel^{-1}\hat k^2\hat u^{\prime\prime},
\label{mompara}
\end{split}
\end{equation}
\begin{equation}
\gamma\taua b=\hat u\hat k\hat B_0+S^{-1}\frac{L^2}{a^2}(\hat b^{\prime\prime}-\hat k^2\hat b).
\label{farapara}
\end{equation}

In equation~(\ref{mompara}) we have retained the higher order derivative (of fourth order) and the term proportional to parallel viscosity. We therefore have neglected terms of order of $k^2\delta^2$ or higher  with respect to $\hat u^{\text{\sc iv}}$, since the velocity gradient scales as   $\sim\delta^{-1}$ in the inner layer, and $k^2\delta^2\ll1$. Parallel viscosity instead introduces a correction of order of $(R/R_\parallel) k^2\delta^2$ with respect to the perpendicular viscous one. This term  should  be retained, as typically $R/R_\parallel\gg1$ in high temperature plasmas.  For instance, in the solar corona $P\approx0.01$ and $P_\parallel\approx10^9$, thus $R/R_\parallel\approx10^{11}$.  

Nevertheless, it is possible to estimate a limit for which parallel viscosity effects  are negligible:  the  fastest growing mode in the inviscid case has both $k\sim S^{-1/4}$ and $\delta\sim S^{-1/4}$~\citep{lou_2013}, so that $(R/R_\parallel) k^2\delta^2\ll~1$ if $R/R_\parallel\ll S$. Such a condition is quite satisfied  for realistic Lundquist numbers  $S\approx10^{12}-10^{14}$. 

\begin{figure}[htbp]
\begin{center}
\includegraphics[width=0.45\textwidth]{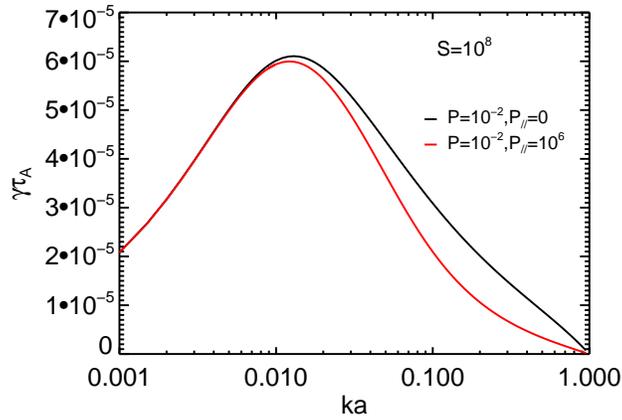}
\caption{Dispersion relations obtained from equations~(\ref{mompara})--(\ref{farapara}) for $S=10^8$, $P=10^{-2}$, $P_\parallel=10^6$ (red) and $P_\parallel=0$ (black). }
\label{S8_para}
\end{center}
\end{figure}

Some effects of parallel viscous terms are shown in Fig.~\ref{S8_para}. Here  we plot dispersion relations obtained from equations~(\ref{mompara})--(\ref{farapara}) for $S=10^8$ and negligible perpendicular viscosity ($P=10^{-2}$), and we compare the growth rates in the case of zero parallel viscosity, $P_\parallel=0$, with those having a large parallel viscosity, $P_\parallel=10^6$ (corresponding to $R/R_\parallel=10^8$). As can be seen, parallel viscous effects are stronger at large wave vectors, and negligible near the fastest growing mode and below.

\section{Discussion:  collapsing current sheets at high Prandtl numbers}
\label{collapsing}

In Section II we described the main properties of the classic visco-resistive tearing instability. We come now to the question of what role viscosity might play in natural systems where current sheets are the outcome of dynamical processes leading to the formation of thin layers.  Following~\citep{pucci_2014}, we therefore study what happens when the current sheet thickness $a$ is allowed to vary. In this case the relevant  unit to define a clock to measure the rapidity of  energy release due to reconnection  is a macroscopic length $L$, that we associate with the length of the sheet. In this way, the aspect ratio $L/a$ is introduced in equations~(\ref{mom})--(\ref{fara}) as a parameter to quantify the contraction of the equilibrium magnetic field.  

Before showing numerical results for unstable modes at arbitrary $L/a$, some considerations are worthwhile.  We found, in the previous Section, where we set $a=L$, that the fastest growing mode has a growth rate which tends to $\gamma_m\tau_{\textsc a}\propto{S}^{-1/2}{P}^{-1/4}$ for both $P\gg1$ and $S\gg1$.  Along the same lines of Ref.~\citep{pucci_2014}, one can redefine  time scales by normalizing them with ${L}$ (see also eq.~(\ref{norm})). Likewise, we find that  for large Prandtl numbers  the maximum growth rate  scales as
\begin{equation}
\gamma_m\tau_{\textsc a}\propto S^{-1/2}\,P^{-1/4}(a/L)^{-3/2}, 
\label{scaling}
\end{equation}
where the constant of proportionality approaches $0.62$, provided both $R\gg1$ and $S\gg1$. In the opposite limit, $P\ll1$, we recover the known results of the inviscid case $\gamma_m\tau_{\textsc a}\propto S^{-1/2}(a/L)^{-3/2}$. Similarly, again for $P\gg1$, the wave vector scales as $k_ma\propto S^{-1/8}\,R^{-1/8}(a/L)^{-1/4}$. The reconnective layer of the fastest growing mode scales as $\delta/a\propto(SR)^{-1/8}(a/L)^{-1/4}$, and that of the marginal mode as $\delta/a\propto(SR)^{-1/6}(a/L)^{-1/3}$.

\begin{figure}[htbp]
\begin{center}
\includegraphics[width=0.45\textwidth]{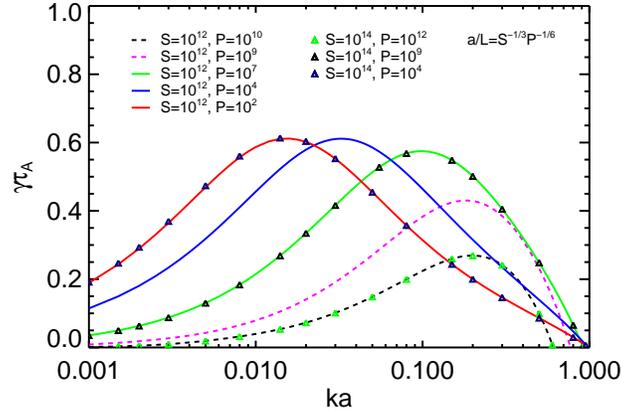}
\caption{Dispersion relation for $a/L=S^{-1/3}P^{-1/6}$ at $S=10^{12}$ (solid lines) and $S=10^{14}$ (triangles) at large Prandtl numbers.}
\label{05}
\end{center}
\end{figure}

Since the maximum growth rate increases for increasing aspect ratio, as shown in  equation~(\ref{scaling}), one can define the critical aspect ratio of the current sheet $L_{\textsc i}/a$ as the one which is unstable  on time scales of order of the Alfv\'en time scale,  thus $L/a_{\textsc i}=S^{1/3}P^{1/6}$.  In Fig.~\ref{05}  we plot the dispersion relation for a current sheet with the critical aspect ratio at realistic Lundquist numbers $S=10^{12}$ (solid lines) and $S=10^{14}$ (triangles), and  large Prandtl numbers.  According to the scalings of $\gamma_m\taua$ and $k_ma$ reported above,  the  dispersion relation does not depend on $S$ (we recall that $P=S/R$), so that curves corresponding to different $S$ superpose exactly, provided  the same Reynolds number is considered. Notice that, for $S\gg1$ and $R\gg1$ the same maximum growth rate $\gamma_m\taua\approx0.62$ is approached. 

 In Fig.~\ref{ga} we show the maximum growth rate versus $a/L$ for   $S=10^{12}$ at  different Prandtl numbers. Almost all the curves have a slope equal to $-3/2$, with the exception of those points at very large $P$ and narrow aspect ratio. This is because the scalings we have inferred are valid as long as a separation of scales between the width of the equilibrium $a$  and the internal reconnective layer $\delta$ is allowed. These constraints cease to be valid when both $a/L\ll1$ and $P\gg1$. 
 For the sake of completeness, we show also in light blue circles the growth rates obtained from equations~(\ref{mompara})--(\ref{farapara}) with parameters relevant to the solar corona and solar flares, $P\approx0.01$ and $P_\parallel\approx10^9$. Growth rates at values of $P\gg1$ are instead appropriate for the solar wind, and the interstellar and intracluster medium (cfr.~table~\ref{table1}).
 
 As shown in Fig.~\ref{ga},  and as can be seen by inspection of equation~(\ref{scaling}), ideal growth rates can now be reached for much larger aspect ratios than in the inviscid case ($P=0$ in the plot), since large viscosity inhibits the growth of the instability. In addition, while in the inviscid case it has been shown that the Sweet-Parker current sheet may not be created naturally, as it turns out that it is much thinner than the critical width of the tearing instability ($a/L_{\textsc {sp}}=S^{-1/2}\ll S^{-1/3}$), now there exists a range of Prandtl numbers for which the viscous Sweet-Parker current sheet width, $a/L_{\textsc {sp}}=S^{-1/2}(1+P)^{1/4}$~\citep{park,Biskamp}, is smaller than, or equal to, the critical width of the visco-tearing instability. To show this point, we represent with asterisks in the plot  the maximum growth rate of  current sheets having  the same inverse aspect ratio of the viscous Sweet-Parker current sheet,    $a/L_{\textsc {sp}}\approx S^{-1/2}P^{1/4}$. In particular, for high Prandtl numbers, the critical aspect ratio equals the aspect ratio of the viscous Sweet-Parker current sheet when $S^{-1/3}P^{-1/6}=S^{-1/2}P^{1/4}$, i.e., for $P=S^{2/5}$. As a consequence, one may expect that for $P\leq S^{2/5}$ tearing instability is disruptive on current sheets thinner than the Sweet-Parker one. The latter, in turn, may  be set as a quasi-stable configuration. 

\section{Conclusions}
\label{conclusions}
In this paper we have analyzed how viscosity influences the tearing mode instability of thin current sheets by spanning from perpendicular Prandtl numbers $P\gg1$ all the way down to $P\ll1$. We have also shown that  large values of parallel Prandtl $P_\parallel$ do not affect growth rates greatly, while the growth of the instability is slowed down if $P\gg1$.

 We have generalized the paper of~\citet{pucci_2014} to show that the asymptotic scaling of the aspect ratio with the Lundquist and (perpendicular) Prandtl number leading to ideal growth rates is $L/a_{\textsc i}=S^{1/3}P^{1/6}$ for $P\gg1$. Large viscosity inhibits the growth of the instability so as to allow for the formation of quasi-stable current sheets thinner with respect to the inviscid case.  This may be important in two respects. 

\begin{figure}[htbp]
\begin{center}
\includegraphics[width=0.45\textwidth]{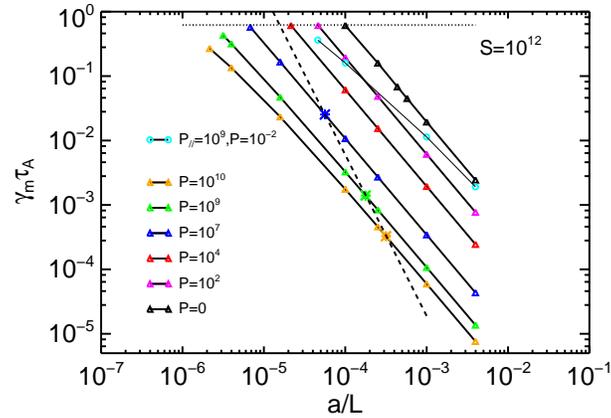}
\caption{Maximum growth rate versus inverse aspect ratio for $S=10^{12}$ at various Prandtl. Dotted line represents the asymptotic growth rate at the critical inverse aspect ratio $a/L=S^{-1/3}P^{-1/6}$ while the dashed line corresponds to the maximum growth rate vs $a/L_{\textsc {sp}}$.}
\label{ga}
\end{center}
\end{figure}

On the one hand we have shown that the viscous Sweet-Parker quasi stationary reconnecting configuration is stable for Prandtl numbers $P\geq S^{2/5}$, for instance, if $S=10^{12}$ then for $P\geq6\times10^4$ (cfr. Fig.~\ref{ga}). As a consequence, viscous stabilization may be important in the solar wind, where reconnection exhausts reminiscent of the Sweet-Parker or Petschek-like configuration are  observed  in regions of relatively large Prandtl numbers  $P\approx3-50$~\citep{gosling_2006, phan_2009}.  Larger values of $P$, of order of  $P\approx10^3-10^5$,  are relevant to the very diluted and hot  intracluster medium, where viscosity may  inhibit reconnection during the dynamo process for  magnetic field amplification in galaxy clusters, and  in protogalaxies~\citep{malyshkin_2002, sheko_2005,lazarian_2011}.

On the other hand, as the stabilizing effect of viscosity allows for the formation of very strong magnetic shears, viscous effects may possibly lead to a smooth transition to kinetic regimes, once the critical width approaches the ion skin depth or the ion Larmor radius.  We are presently working on generalizing the above scalings to kinetic regimes. 
 
\acknowledgements{We wish to thank D. Del Sarto and F. Pegoraro for useful discussions. This research was carried out at the Jet
Propulsion Laboratory, California Institute of Technology, under a contract
with the National Aeronautics and Space Administration. Copyright 2014. All rights reserved. }




\begin{table*}[htdp]
\begin{center}
\begin{tabular}{|cccccc|}
\hline
			&$n$&$T$&$B$&$P$ & $P_\parallel$\\	
\hline
\hline
Solar corona &$10^{9}$&$10^6$&$50$& $10^{-2}$ & $10^9$\\
Solar flares &$10^{10}$&$10^6-10^7$&$100$& $10^{-2}-10^{-1}$ & $10^8-10^{12}$\\
Solar wind &$5$&$(35-23)\times10^4$&$(20-100)\mu$& $3-50$ & $10^{16}-10^{15}$\\
ISM (ionized)&$0.2-0.006$&$10^4-10^6$&$(1-10)\mu$& $10-100$ & $10^{11}-10^{20}$\\
Intracluster medium &$10^{-3}$&$10^8$&$(0.1-1)\mu$& $10^5-10^3$ & $10^{29}$\\
\hline
\end{tabular}
\end{center}
\caption{Some examples of magnetized plasmas in space  and their order of magnitude parameters in cgs units.}
\label{table1}
\end{table*}%


\begin{thebibliography}{}

\bibitem[{\it Biskamp(1993)}]{Biskamp}  Biskamp, D.,  Nonlinear Magnetohydrodynamics, Cambridge Monographs on Plasma Physics 1, Cambridge University Press (1993).

\bibitem[{\it Bondeson and Sobel(1984)}]{bondeson} Bondeson, M., and Sobel, J.R., Energy balance of the collisional tearing mode, Physics of Fluids, {\bf 27}, 2028 (1984). 

\bibitem[{\it Braginskii(1965)}]{brag}  Braginskii, S.I.,  Transport Processes in a Plasma,  Reviews of Plasma Physics,  {\bf 1}, 205 (1965). 


\bibitem[{\it Cassak et al.(2005)}]{cassak_prl_2005}  Cassak, P.A., Shay, M.A., and Drake, J.F., Catastrophe Model for Fast Magnetic Reconnection Onset, Phys. Rev. Lett. {\bf 95}, 235002 (2005).

\bibitem[{\it Cassak and Drake(2013)}]{cassak_pop_2013}  Cassak, P.A., and Drake, J.F., On the phase diagrams of magnetic reconnection, Phys. Plasmas, {\bf 20}, 061207 (2013).

\bibitem[{\it Cerri et al. (2013)}]{cerri_2013} Cerri, S.S., Henri, P., Califano, F. et al., Extend fluid models: pressure tensor effects and equilibria,  Phys. Plasmas {\bf 20}, 112112 (2013)

\bibitem[{\it Dobrowolny et al.(1983)}]{dobro}  Dobrowolny, M., Veltri, P., and Mangeney, A., Dissipative instabilities of magnetic neutral layers with velocity shears, Journal of Plasmas Physics, {\bf 29}, 393 (1983). 

\bibitem[{\it Furth et al.(1963)}]{FKR}  Furth, H.P.,  Killeen, J., and  Rosenbluth, M.N., Finite Resistivity Instabilities of a Sheet Pinch, Physics of Fluids, {\bf 20}, 459 (1963).

\bibitem[{\it Gosling et al.(2005)}]{gosling_2006}  Gosling, J.T., Eriksson, S., and Schwenn, R., Petschek-type magnetic reconnection exhaust in the solar wind well inside 1 AU: Helios, Journal of Geophysical Research, {\bf 111}, A10102 (2006).

\bibitem[{\it Grasso et al.(2008)}]{grasso_pop_2008}  Grasso, D., Hastie, R.J., Porcelli, F., and Tebaldi, C., Physics of Plasmas, {\bf 15}, 072113 (2008).

\bibitem[{\it Lentini and Pereira(1974)}]{lentini} Lentini, M., and Pereyra, V., A variable order finite difference method for nonlinear multipoint boundary value problems, Math. Comp., {\bf 28}, 9811004 (1974). 

\bibitem[{\it Lazarian and Brunetti(2011)}]{lazarian_2011} Lazarian, A., and Brunetti, G., Turbulence, reconnection and cosmic rays in galaxy clusters, Mem. S.A.It., {\bf 82}, 636 (2011).

\bibitem[{\it Loureiro et al. (2007)}]{lou_2007} Loureiro, N.F., Schekochihin, A.A., Cowley, S.C., Instability of current sheets and formation of plasmoid chains, Physics of Plasmas, {\bf 14}, 100703  (2007). 

\bibitem[{\it Loureiro et al.(2012)}]{lou_2012} Loureiro, N.F., Samtaney, R., Schekochihin, A.A., and Uzdensky, D., A., Magnetic reconnection and stochastic plasmoid chains in high-Lundquist number
plasmas, Physics of Plasmas, {\bf 19}, 042303 (2012). 

\bibitem[{\it Loureiro et al. (2013)}]{lou_2013} Loureiro, N.F., Schekochihin, A.A., Uzdensky, D.A., Plasmoid and Kelvin-Helmholtz instabilities in Sweet-Parker current sheets, Physical Review E, {\bf879}, 013102  (2013). 

\bibitem[{\it Malara and Velli(1996)}]{malara_96} Malara, F., and Velli, M., Phys. Plasmas, {\bf 3}, 4427 (1996).

\bibitem[{\it Malyshkin and Kulsrud(2002)}]{malyshkin_2002}  Malyshkin, M., and Kulsrud, R.M., Magnetized turbulent dynamos in protogalaxies, The Astrophysical Journal, {\bf 571}, 619 (2002).

\bibitem[{\it Militello et al.(2011)}]{militello_pop_2011}  Militello, F., Borgogno, D., Grasso, D., Marchetto, C., and Ottaviani, M., Phys. Plasmas, {\bf 18}, 112108 (2011).

\bibitem[{\it Ofman et al.(1991)}]{ofman} Ofman, L., Chen, K.L., and Morrison, P.J., Physics of Fluids B {\bf 6}, 1364 (1991).

\bibitem[{\it Park et al.(1984)}]{park} Park, W., Monticello, D. A., and White, R.B., Reconnection rates of magnetic fields including the effects of viscosity, Phys. Fluids {\bf 27}, 137 (1984). 

\bibitem[{\it Phan et al.(2009)}]{phan_2009}  Phan,T.D., Gosling, J.T.,  and  Davis, M.S., Prevalence of extended reconnection X-lines in the solar wind at 1 AU, Geophys. Res. Letters, {\bf 36}, L09108 (2009).

\bibitem[{\it Porcelli(1987)}]{porcelli_1987} Porcelli, F., Viscous resistive magnetic reconnection, Phys. Fluids {\bf 30}, 1734 (1987). 

\bibitem[{\it Pucci and Velli(2014)}]{pucci_2014} Pucci,  F., and Velli, M., Reconnection of quasi-singular current sheets: the "ideal" tearing mode, The Astrophysical Journal Letters, {\bf 780} (2014). 

\bibitem[{\it Rappazzo and Parker(2013)}]{rappa_2013} Rappazzo, A.F., Parker, E.N., Current sheets formation in tangled coronal magnetic fields, The Astrophysical Journal Letters, {\bf 773}, L2 (2013). 

\bibitem[{\it Schekochihin et al.(2005)}]{sheko_2005}  Schekochihin, A.A.,  Cowley,  S.C.,  Kulsrud, R.M.,    Hammett, G.W.,  and Sharma, P., Plasma instabilities and magnetic field growth in cluster of galaxies, The Astrophysical Journal, {\bf 629}, 139 (2005).

\bibitem[{\it Velli and Hood(1989)}]{Velli_1989} Velli, M., and Hood, W., Solar Physics, {\bf 119}, 107 (1974). 

\end{thebibliography}
\end{document}